\documentclass[unnumsec,webpdf,contemporary,large]{oup-authoring-template}%

\usepackage{isomath}
\usepackage{amsmath,amsthm}
\usepackage{amsbsy}
\usepackage{amssymb}
\usepackage{amscd}
\usepackage{amsfonts}
\usepackage{stmaryrd}
\usepackage{euscript}
\usepackage[utf8]{inputenc}
\usepackage[T1]{fontenc}
\usepackage{newtxtext} 
\everymath{\displaystyle}

\usepackage{hyperref}

\usepackage{graphicx}
\usepackage{boxedminipage}
\usepackage{calc}
\graphicspath{ {media/} }

\usepackage{subcaption}

\usepackage{orcidlink}
\usepackage{siunitx}

\renewcommand{\figurename}{Figure}

\renewcommand{\thetable}{\arabic{table}}

\usepackage{isomath}
\usepackage{amsmath}
\usepackage{amssymb}
\usepackage{amscd}
\usepackage{amsfonts}

\newcommand{\bfbeta}{\mathbold {\beta}}

\newcommand{\dm}{\ \mathrm{d}}

\newcommand{\bfe}{{\mathbold e}}

\newcommand{\bfp}{{\mathbold p}}

\newcommand{\bfr}{{\mathbold r}}

\newcommand{\bfu}{{\mathbold u}}

\newcommand{\bfx}{{\mathbold x}}

\newcommand{\bfE}{{\mathbold E}}

\newcommand{\bfI}{{\mathbold I}}

\graphicspath{{Fig/}}

\theoremstyle{thmstyleone}%
\theoremstyle{thmstyletwo}%
\theoremstyle{thmstylethree}%

\begin{document}

\journaltitle{PNAS Nexus}
\DOI{https://doi.org/10.1093/pnasnexus/pgaf281}
\copyrightyear{2025}
\pubyear{2025}
\access{\ }
\appnotes{\ }

\firstpage{1}


\title[Dipolar Self-Interactions Drive Polymer Chain Collapse]
{Nonlocal Dipolar Self-Interactions Drive Polymer Chain Collapse in Electric Fields}


\author[a]{Pratik Khandagale}
\author[b, c]{Gal deBotton}
\author[d]{Timothy Breitzman}
\author[a, e, f]{Carmel Majidi}
\author[e, a, g, $\ast$]{Kaushik Dayal}

\authormark{Khandagale et al.}

\address[a]{\orgdiv{Department of Mechanical Engineering}, \orgname{Carnegie Mellon University}, \orgaddress{\street{5000 Forbes Ave, Pittsburgh}, \postcode{15213}, \state{PA}, \country{USA}}}
\address[b]{\orgdiv{Department of Mechanical Engineering}, \orgname{Ben Gurion University}, \orgaddress{\street{1 Ben-Gurion Blvd, Beer-Sheva}, \postcode{8410501}, \country{Israel}}}
\address[c]{\orgdiv{Department of Biomedical Engineering}, \orgname{Ben Gurion University}, \orgaddress{\street{1 Ben-Gurion Blvd, Beer-Sheva}, \postcode{8410501}, \country{Israel}}}
\address[d]{\orgdiv{Materials and Manufacturing Directorate}, \orgname{Air Force Research Laboratory}, \orgaddress{\street{Wright-Patterson Air Force Base}, \postcode{45433}, \state{OH}, \country{USA}}}
\address[e]{\orgdiv{Department of Civil and Environmental Engineering}, \orgname{Carnegie Mellon University}, \orgaddress{\street{5000 Forbes Ave, Pittsburgh}, \postcode{15213}, \state{PA}, \country{USA}}}
\address[f]{\orgdiv{Department of Materials Science and Engineering}, \orgname{Carnegie Mellon University}, \orgaddress{\street{5000 Forbes Ave, Pittsburgh}, \postcode{15213}, \state{PA}, \country{USA}}}
\address[g]{\orgdiv{Center for Nonlinear Analysis, Department of Mathematical Sciences}, \orgname{Carnegie Mellon University}, \orgaddress{\street{5000 Forbes Ave, Pittsburgh}, \postcode{15213}, \state{PA}, \country{USA}}}

\corresp[$\ast$]{To whom correspondence should be addressed: \href{email:Kaushik.Dayal@cmu.edu}{Kaushik.Dayal@cmu.edu}}


\received{Date}{0}{Year}
\accepted{Date}{0}{Year}

\boxedtext{
    Dielectric polymers provide actuation, sensing, and energy harvesting capabilities in emerging soft matter-based technologies.
We report for the first time that a dielectric polymer chain collapses in a high electric field and that it can be controlled by the applied field orientation and mechanical stretching.
Our novel theoretical approach enables deeper insights into the complex effects of nonlocal interactions, which are challenging to account for.
For experiments, embedding collapse-prone dielectric polymers within material architectures creates opportunities for rapid actuation and sensing technologies using electrical stimuli.
Our findings provide a pathway to discover the physics behind instabilities and electrical breakdown in dielectric polymers that critically limit their real-world applications.
}

\abstract{
We report that a dielectric polymer chain, constrained at both ends, sharply collapses when exposed to a high electric field. The chain collapse is driven by nonlocal dipolar interactions and anisotropic polarization of monomers, a characteristic of real polymers that prior theories were unable to incorporate. Once collapsed, a large number of chain monomers accumulate at the center location between the chain ends, locally increasing the electric field and polarization by orders of magnitude. The chain collapse is sensitive to the orientation of the applied electric field and chain stretch. Our findings not only offer new ways for rapid actuation and sensing but also provide a pathway to discover the critical physics behind instabilities and electrical breakdown in dielectric polymers.}


\maketitle

\section{Introduction}

\begin{figure*}[!t]
\centering
{\includegraphics[width=\linewidth]{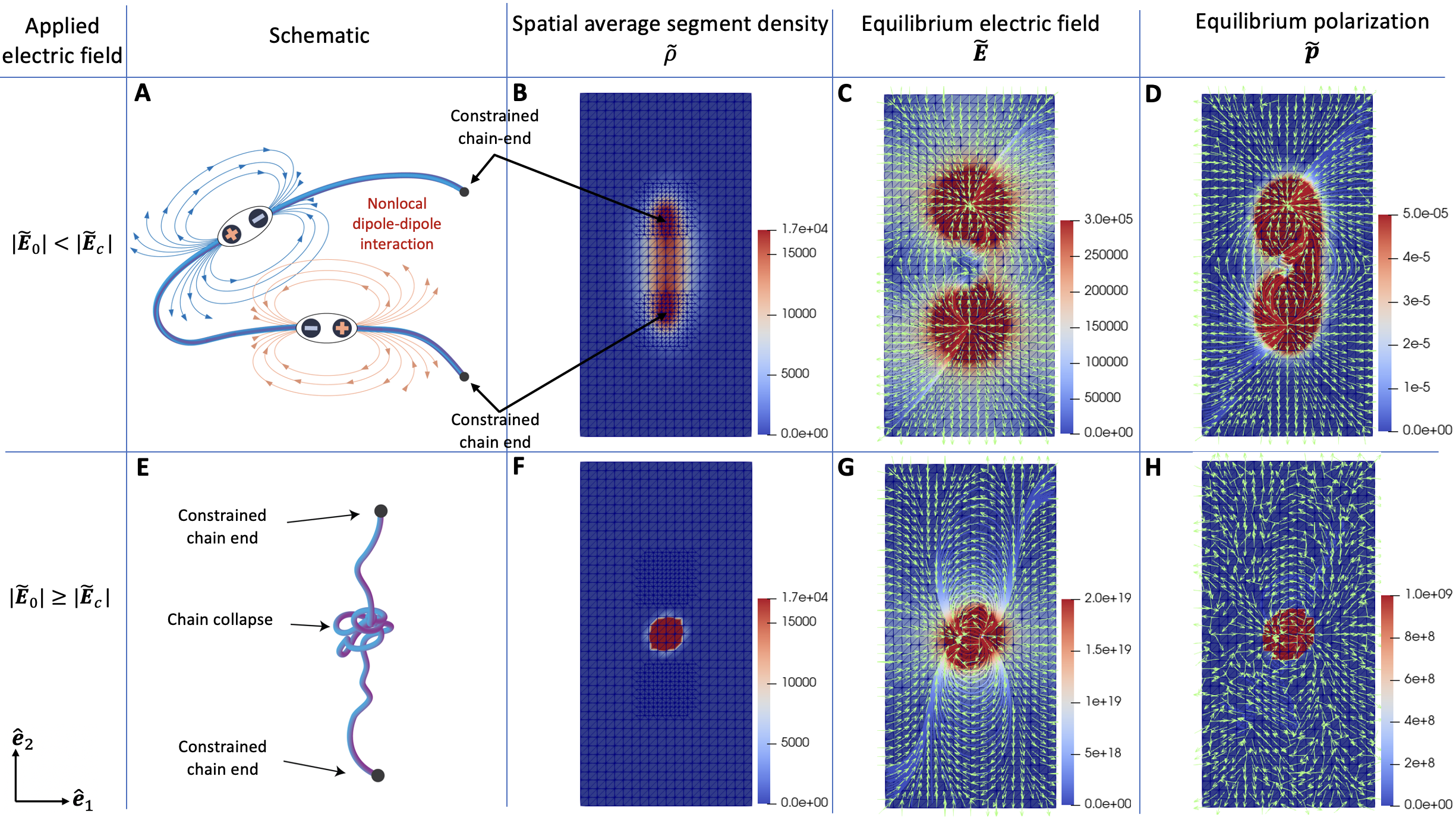}}
\caption{Chain collapse phenomenon of a dielectric polymer chain with constrained chain ends in a high electric field. 
A) 
The dielectric polymer chain is in a non-collapsed state when the strength of the applied electric field $(|\tilde{\bfE}_0|)$ is less than a critical value for chain collapse $(|\tilde{\bfE_c}|)$. 
The applied electric field induces electric dipoles in the polymer segments, where each segment represents a set of connected monomers along the chain contour in a coarse-grained setting. In addition to the interaction between the electric field and dipoles, the dipoles themselves interact with each other nonlocally. 
B) Spatial average segment density $\tilde{\rho}$, C) electric field $\tilde{\bfE}$, and D)  polarization $\tilde{\bfp}$ for the non-collapsed chain (here $\tilde{\bfE}_0= 620 \hat{\bfe}_2$). 
E) 
When the strength of the applied electric field is higher than a critical value (i.e., $|\tilde{\bfE}_0| \geq |\tilde{\bfE}_c|)$, the nonlocal dipole-dipole interactions lead to a chain collapse resulting in the accumulation of a large number of polymer segments near the center location between the constrained chain ends. 
F) Spatial average segment density $\tilde{\rho}$, G)  electric field $\tilde{\bfE}$, and H)  polarization $\tilde{\bfp}$ for the collapsed chain (here $\tilde{\bfE}_0= 625 \hat{\bfe}_2$).  
In the collapsed state, the electric field and polarization have significantly altered spatial distribution, and they are orders of magnitude higher when compared to the non-collapsed state (as indicated by different color scales). The chain is under $30\%$ stretch.
}\label{fig:intro}
\end{figure*}

Polarization of dielectric polymers in an external electric field provides sensing, actuation, and energy-harvesting functionalities in emerging technologies such as soft robotics 
\cite{shian2015dielectric, 
rich2018untethered, 
ji2019autonomous,  gu2017survey, thuruthel2019soft, chan2012development},
stretchable electronics \cite{rogers2010materials, xu2013stretchable, grasinger2021flexoelectricity, deng2014electrets, zhao2021modeling, chen2021interplay, ahmadpoor2015flexoelectricity, mathew2024electro, zolfaghari2020network, gurnani2024ai},
and artificial muscles 
\cite{mirvakili2018artificial, mirfakhrai2007polymer, pelrine2002dielectric, mu2019sheath, chen2024soft}.
This versatile multi-functionality of dielectric elastomers stems from the fundamental electro-mechanical response of a single dielectric polymer chain in an external electric field. 
Here, we report for the first time that when a dielectric polymer chain constrained at its two ends and having an anisotropic polarization response is subjected to a high external electric field, it undergoes a sharp conformational change: a very large fraction of the monomers accumulate at the center between the constrained ends (Fig. \ref{fig:intro}E), which we refer to as \textit{chain collapse}. 
We highlight that chain collapse occurs only when the dielectric polymer chain has an anisotropic electrical response, as is characteristic of many real polarizable monomers. 
Existing polymer field theories for dielectric polymers (e.g., \cite{martin2016statistical, budkov2015communication, budkov2016polarizable}) are restricted to isotropic or perturbative weak anisotropic polarization response, and cannot capture this behavior.
The anisotropic response introduces essential physics: it couples mechanical deformation and the electric fields in a highly nonlinear way.
Anisotropic monomers pay an energetic price if they are not aligned along the electric field, setting up a competition between entropic polymer elasticity and energetic dielectric response; 
whereas for an isotropic monomer, there is no electrostatic coupling to monomer orientation.

Large conformational changes in polymers with external stimuli, in general, provide us with a promising way to not only exploit them in modern engineering applications but also discover novel soft matter-based technologies. 
Phase transitions in polymers leading to significant conformational changes have been discovered and studied in the past 
\cite{brilliantov1998chain, tom2016mechanism, jia2022dipole, hsiao2006salt, netz2003nonequilibrium, kang2015effects, philip2002stretching, long1996simultaneous, jiao2014understanding, mathew2025active,  locatelli2021activity, monari1999sequence, baiesi2006scaling, zhou2006collapse, he2012partial}.
For example, when a bad solvent (the one that mediates attractive interactions among polymer monomers) replaces a good solvent (the one that mediates repulsive interactions among polymer monomers) surrounding the polymer chain, 
the chain collapses from a coil configuration to a globule phase 
\cite{swislow1980coil, maki2014chain, dua1999polymer,  de1975collapse, loh2008collapse}.
A polyelectrolyte polymer, such as DNA, whose polymer segments carry permanent dipoles, collapses in the presence of a high external electric field \cite{radhakrishnan2021collapse, zhou2011collapse, zhou2015collapse, tang2011compression}.
This transition is driven by an increase in attractive dipole-dipole interactions among DNA chain segments. A polyelectrolyte gel, a network of polyelectrolyte polymer chains with a solvent, collapses in an external electric field through discrete and reversible volume change \cite{tanaka1982collapse}.

Dielectric polymer chains, on the other hand, possess induced electric dipoles in the presence of an external electric field. Its physics is driven by the precise nature of polarization response and both local as well as nonlocal electrostatic interactions. The local interactions consist of the interactions between the external electric field and induced dipoles, and the nonlocal interactions are the interactions among induced dipoles (Fig. \ref{fig:intro}A), which are challenging to account for. 
Theories developed in the past have investigated the conformational changes in dielectric polymers. 
Flory-type phenomenological theory predicted that an external electric field induces a globule-to-coil transition for a dielectric polymer chain in a bad solvent 
\cite{budkov2016polarizable}. 
However, this model assumed an isotropic dielectric response, a sufficiently large polymer volume, and the random phase approximation, while accounting for many-body dipole-dipole interactions. 
The statistical theory for dielectric polymer proposed in \cite{martin2016statistical} accounts for dipole-dipole interactions, but it only applies to dielectric polymers having isotropic polarization response. 
The limitations of these existing theories for dielectric polymers to account for nonlocal dipolar interactions as well as anisotropic polarization response have been overcome by a statistical field theory model for dielectric polymers in \cite{khandagale2024statistical}. This theory paved the way to advance the applicability of dielectric polymers in real-world scientific and engineering applications. 
We adopt this theory and solve it numerically to show the chain collapse phenomenon for a dielectric polymer chain in a high electric field. 
The theory adopted in \cite{khandagale2024statistical} is based on statistical mechanics, which enables us to include polymer molecular details, a self-consistent field theory that enables us to account for the nonlocal dipole-dipole interactions, and electrostatics.

In this report, we show that nonlocal dipole-dipole interactions, as well as the anisotropic polarization response of monomers, are the necessary factors for the chain collapse of the dielectric polymer in a high electric field. 
Chain collapse occurs when the strength of the applied electric field is greater than a critical value.  
A higher applied electric field would induce higher chain polarization, which implies higher dipole-dipole attractive interactions. These enhanced attractive interactions would force the mobilization of chain segments to accumulate and collapse together, providing the physical rationale for the observed chain collapse phenomenon.   
We show that the critical electric field for chain collapse is sensitive to the orientation of the applied electric field as well as the stretching of the chain. 

The chain collapse phenomenon reported here provides insights into the critical physics that arises at a polymer chain length scale due to the microscopic nonlocal dipole-dipole interactions and molecular anisotropy in the polarization response of chain monomers. It also opens up new avenues to exploit chain collapse instability for novel applications of dielectric polymers under external electric field stimuli, such as rapid actuation, sensing, and its control using mechanical deformation and orientation of the applied electric field.

\section{Theoretical formulation}

We consider the polymer chain to be flexible, with $N$ segments, where each segment represents a set of connected monomers along the polymer chain backbone in a coarse-grained setting. Each segment is considered inextensible, which makes the whole chain inextensible. 
Each segment has a length $a=L_c/N$, where $L_c$ is the total contour length of the chain. 

\subsection{Statistical field theory of a polymer chain}
\begin{figure}
\centering
{\includegraphics[width=\linewidth]{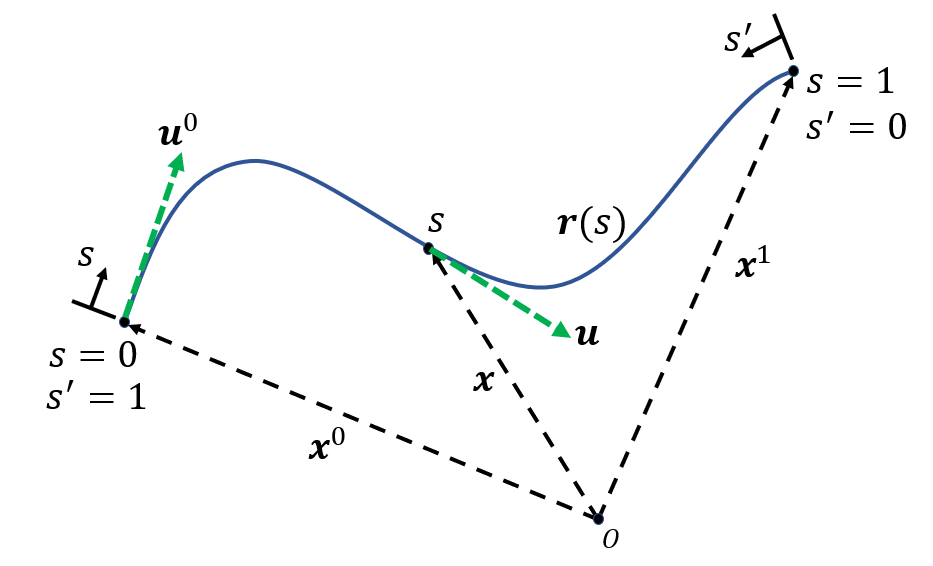}}
\caption{Description of polymer chain as a continuous curve.}\label{fig:single_worm_like_chain}
\end{figure}
We use the statistical field theory formulation for dielectric polymer chains in an external electric field, adopted from \cite{khandagale2024statistical}.
The polymer chain is described using a worm-like chain (WLC) model \cite{fredrickson2006equilibrium, spakowitz2004exact} with a small persistence length $\lambda$, where $\lambda$ denotes the distance along the polymer chain contour over which the orientational correlation decays. 
Parameter $\lambda$ depends on the molecular structure of the polymer chain.  
The flexibility of the polymer chain is determined by the ratio $\lambda/L_c$, where $\lambda/L_c \ll 1$ corresponds to a very flexible polymer chain, whereas $\lambda/L_c \gg 1$ models a rigid rod-like polymer chain.  

A polymer chain is represented as a thermally fluctuating continuous 3-d space curve $\bfr(s)$ in a coarse-grained setting (Fig. \ref{fig:single_worm_like_chain}). Here $\bfx=\bfr(s)$ denotes the spatial position of a polymer segment at the chain contour coordinate $s$ that varies along the chain contour. The coordinate $s$ varies as $0 \leq s \leq 1$ and is nondimensionalized with $L_c$. 
The chain segment orientation at any $s$ is given by $\bfu=\frac{1}{L_c}\frac{\dm\bfr(s)}{\dm s}$.
Here, we choose $\bfu$ to be a unit vector to ensure that the chain is inextensible. The WLC model has an energy cost for bending, and the total bending energy is given by summing over the chain contour using the harmonic form \cite{fredrickson2006equilibrium}:
\begin{equation}\label{eq:bending_energy}
\begin{split}
	U_b[\bfu] = \frac{\lambda k_B T}{2} \int\limits_{0}^{1} \dm s \Bigg| \frac{\dm \bfu(s)}{\dm s} \Bigg|^2,
\end{split}
\end{equation}
where $\dm \bfu(s)/\dm s$ is the local curvature of the polymer at contour coordinate $s$, and $k_B T$ is the thermal energy scale. 

When the chain is under the influence of a scalar potential $w(\bfx, \bfu)$ that can influence both spatial position $\bfx$ and orientation $\bfu$ of polymer segments, we aim to obtain the equilibrium properties of the polymer chain. 
The equilibrium thermodynamic average of any fluctuating quantity $(\cdot)$ that depends on the fluctuating polymer chain conformation $\bfr(s)$ is obtained as:
\begin{equation}\label{eq:def_stat_avg}
    \langle (\cdot) \rangle = 
    \frac{ \int \mathcal{D} \bfr(s)\, (\cdot)\, \exp\left( -\frac{E[\bfr(s)]}{k_B T} \right) }
         { \int \mathcal{D} \bfr(s)\, \exp\left( -\frac{E[\bfr(s)]}{k_B T} \right) }.
\end{equation}
Here, $E[\bfr(s)]$ is the total energy of the polymer chain in chain conformation $\bfr(s)$, and $\int \mathcal{D} \bfr(s)$ denotes the functional integration over all possible chain conformations $\bfr(s)$ that the chain can possess.
The denominator in \eqref{eq:def_stat_avg} is the partition function, a fundamental quantity in statistical mechanics used to account for the entropy and obtain equilibrium
thermodynamic properties of the system.  
For example, the Helmholtz free energy of the WLC in the external potential $w$ is obtained using its partition function denoted by $Q[w]$ as $-k_B T \log Q[w]$.  
The partition function is a moment generating function that can be used to obtain various moments (the expected values) of the fluctuating quantities of the system. 

The analytical expression for the WLC partition function $Q[w]$ is given by:
\begin{equation}\label{eq:Q}
    Q[w]= \frac{1}{4 \pi V } \int \dm \bfx \int \dm \bfu \ q(\bfx, \bfu, \bfx^0, \bfu^0, s) \ q^{*}(\bfx, -\bfu, \bfx^{1}, \bfu^{1}, 1-s).
\end{equation}  
Here, $\bfx^0$, $\bfu^0$ denote the position and orientation vectors, respectively, at the chain end $s=0$, and $\bfx^{1}$, $\bfu^{1}$ denote the position and orientation vectors, respectively, at the chain end $s=1$. $V = Na^3$ is the total volume of the polymer chain in 3-d spatial dimension (and one would use $V=Na^2$ for 2-d).
The integration in \eqref{eq:Q} is performed over the combined space composed of the segment position $\bfx$ and the unit sphere space spanned by segment orientation $\bfu$. 
The quantities $q$ and $q^*$ represent the propagation of correlations in segment position $\bfx$ and orientation $\bfu$ for
WLC under an external potential $w$.
Specifically, $q(\bfx, \bfu, \bfx^0, \bfu^0, s)$ represents the probability density 
that the chain contour coordinate $s$ has spatial position $\bfx$ and orientation $\bfu$, where the initial end of the chain fragment starts from $s = 0$ having spatial position $\bfx^0$ and chain orientation $\bfu^0$. 
The quantities $q$ and $q^{*}$ are also known as partial partition functions, where $q$ corresponds to the chain fragment starting from chain contour coordinate $s=0$ to $s$, whereas $q^{*}$ corresponds to the other chain fragment starting from the opposite chain end with chain contour coordinate $s=1$ to $s$ (Fig. \ref{fig:single_worm_like_chain}).
These quantities are governed by the partial differential equations (PDEs) below:
\begin{align}
	\label{eq:q_PDE}
	\frac{\partial q}{\partial s} &= -w(\bfx, \bfu) q -L_c \bfu \cdot \nabla_{\bfx} q + \frac{L_c}{2 \lambda} \nabla^2_{\bfu}q,
	\\
	\label{eq:q_star_PDE}
	\frac{\partial q^{*}}{\partial s'} &= -w(\bfx, \bfu) q^{*} - L_c \bfu \cdot \nabla_{\bfx} q^{*} + \frac{L_c}{2 \lambda} \nabla^2_{\bfu}q^{*}.
\end{align}
Here, the chain contour coordinate $s' = 1-s$ varies in the opposite direction as $s$ (Fig. \ref{fig:single_worm_like_chain}). 
The initial conditions for the above PDEs are:
\begin{align}
	\label{eq:q_ini_condition}
	q(\bfx, \bfu, \bfx^0, \bfu^0, s)\Big|_{s=0} &= V \ \delta (\bfx- \bfx^0),
	\\
	\label{eq:q_star_ini_condition}
	q^{*}(\bfx, -\bfu, \bfx^{1}, \bfu^{1}, s')\Big|_{s'=0} &= V \ \delta (\bfx- \bfx^{1}).
\end{align}
The initial conditions in \eqref{eq:q_ini_condition} and \eqref{eq:q_star_ini_condition} specify the physical constraint that the chain ends are fixed at $\bfx^0$ and $\bfx^{1}$. 
We have not constrained $\bfu^0$ and $\bfu^{1}$, the chain segment orientations at the two chain ends.  
%

The linear PDEs in \eqref{eq:q_PDE} and \eqref{eq:q_star_PDE} for $q$ and $q^*$ can be viewed as stochastic differential equations (specifically, as Fokker-Planck equations) that govern the propagation of correlations in segment position $\bfx$ and orientation $\bfu$ for WLC under an external potential $w$.
These PDEs are derived using the stochastic process analogy of a polymer chain and a recursive relation for the chain partition function based on the Markov property of the polymer chain, while accounting for the harmonic bending energy of WLC stated in \eqref{eq:bending_energy} (Section 2.5 in \cite{fredrickson2006equilibrium}). 
The left-hand side of the PDE for $q$ in \eqref{eq:q_PDE} represents the evolution of $q$ over the chain contour coordinate $s$.
The first term on the right-hand side of \eqref{eq:q_PDE} accounts for the external potential $w$. The second term, $-L_c \bfu \cdot \nabla_{\bfx} q$, can be viewed as a drift term that dictates the average direction of the evolution of $q$. The third term, $\frac{L_c}{2 \lambda} \nabla^2_{\bfu} q$ captures the randomness in the evolution of $q$; specifically, the rotational diffusion operator $\nabla^2_{\bfu}$ generates diffusive motion on the unit sphere, and 
the parameter $\frac{L_c}{2 \lambda}$ can be viewed as a rotational diffusion coefficient. Consistent with physical understanding, the diffusion term that is responsible for the randomness in the propagation of $q$ and $q^*$ is dominant for a flexible polymer chain $(\lambda/L_c \ll 1)$ compared to a stiff polymer chain $(\lambda/L_c \gg 1)$.

The density operator $\hat{\rho}(\bfx, \bfu)$ of the WLC, that depends on the fluctuating chain conformation $\bfr(s)$, is defined as \cite{fredrickson2006equilibrium}:
\begin{equation}\label{eq:rho_def}
    \hat{\rho}(\bfx, \bfu):= \int\limits_{0}^{1} \dm s \ \delta(\bfx- \bfr(s)) \ \delta \left(\bfu - \frac{1}{L_c}\frac{\dm\bfr(s)}{\dm s} \right) \delta(|\bfu|-1).
\end{equation}
Here, the kinematic definitions of $\bfx$ and $\bfu$ are embedded using the Dirac measures as constraints.  
The thermodynamically-averaged segment density $\langle \hat{\rho}(\bfx, \bfu) \rangle$ is then obtained using the definition of statistical average in \eqref{eq:def_stat_avg}.
This thermodynamic averaging can be derived in terms of the partition function to have the convenient form \cite{fredrickson2006equilibrium}:
\begin{equation}\label{eq:rho_avg}
\begin{split}
	& \langle  \hat{\rho}(\bfx, \bfu) \rangle  \\
    & = \frac{1}{ 4 \pi V Q[w]} 
    \int\limits_{0}^{1} \dm s \ q(\bfx, \bfu, \bfx^0, \bfu^0, s) q^{*}(\bfx, -\bfu, \bfx^{1}, \bfu^{1}, 1-s).
\end{split}
\end{equation}
 
\subsection{Spatial dipole distribution}






To obtain the equilibrium chain polarization in a statistical field theory framework, we define a polarization operator $\hat{\bfp}(\bfx, \bfu)$ that depends on the fluctuating chain conformation $\bfr(s)$ as:
\begin{equation}\label{eq:polarization_operator_x_u}
\begin{split}
   &  \hat{\bfp}(\bfx, \bfu)  := 4 \pi V  \\
    & \int\limits_{0}^{1} \dm s \ \bfp_{seg}(\bfx, \bfu)  \ \delta(\bfx- \bfr(s)) \  \delta \left(\bfu- \frac{1}{L_c}\frac{\dm\bfr(s)}{\dm s} \right) \delta(|\bfu|-1) ,
\end{split}
\end{equation}
where $\bfp_{seg}(\bfx, \bfu)$ is the polarization response function of the chain segment.
 We define $\langle \hat{\bfp}(\bfx, \bfu) \rangle $ as the thermodynamically-averaged polarization. By taking a statistical average on both sides of \eqref{eq:polarization_operator_x_u} and using \eqref{eq:rho_def}-\eqref{eq:rho_avg}, we obtain $\langle \hat{\bfp}(\bfx, \bfu) \rangle $ as: 
\begin{equation}\label{eq:avg_polarization_x_u_appendix}
\begin{split}
    & \langle \hat{\bfp}(\bfx, \bfu) \rangle 
    = 
    4 \pi V \bfp_{seg}(\bfx, \bfu) \left\langle  \hat{\rho}(\bfx, \bfu)  \right\rangle, 
\end{split}
\end{equation} 
By averaging $\langle \hat{\bfp}(\bfx, \bfu) \rangle $ over segment orientation space $\bfu$, we obtain polarization at the spatial location $\bfx$, denoted by $\bfp(\bfx)$ as:
\begin{equation}\label{eq:avg_polarization_x}
  \bfp(\bfx) := \frac{1}{4 \pi} \int \dm \bfu \
            \langle \hat{\bfp}(\bfx, \bfu) \rangle  = V \int \dm \bfu \
            \bfp_{seg}(\bfx, \bfu)  \langle  \hat{\rho}(\bfx, \bfu) \rangle .
\end{equation} 
Quantity $\bfp(\bfx)$ will enter in the electrostatic equation \eqref{eq:electrostatic} in the bound charge density.

\subsection{Electrostatic  interactions}\label{sec:self_consistent_formulation}

We obtain the electric field $\bfE(\bfx)$ from polarization $\bfp(\bfx)$ in \eqref{eq:avg_polarization_x} using the electrostatic equation (see Supplementary Material Sec. 3 for derivation):
\begin{equation}\label{eq:electrostatic}
    - \epsilon_0 \nabla_{\bfx}^2 \phi(\bfx)= -\alpha \  \nabla_{\bfx} \cdot \bfp \text{ on } \Omega, \quad \text{ given } \phi(\bfx)= -\bfE_0\cdot \bfx \text{ on } \partial\Omega,
\end{equation} 
where $\phi(\bfx)$ is the electric potential throughout the spatial domain $\Omega$; $-\nabla_{\bfx} \cdot \bfp$ is the bound charge density; and $\bfE_0$ is the applied average electric field on the domain boundary $\partial\Omega$. 
Note that we introduced a scalar parameter $\alpha \in [0,1]$ in \eqref{eq:electrostatic} to scale the dipole-dipole interactions that are being accounted for. $\alpha=0$ implies no dipole-dipole interactions, whereas $\alpha=1$ implies accounting for full dipole-dipole interactions. 
Since the charge distribution described through $\bfp(\bfx)$ does not involve singular dipole distributions, the local electrostatic PDE in \eqref{eq:electrostatic} is numerically tractable. 
This PDE also allows us to apply realistic boundary conditions used in lab experiments without having to compute the Green's function for a given domain geometry, e.g, the specified electric potential on the domain boundary corresponding to a given far-field applied electric field. 
The electric field $\bfE(\bfx)$ within the domain is dictated by electrostatic interactions between the applied electric field and the induced dipoles, as well as the nonlocal dipole-dipole interactions.
The electrostatic PDE in \eqref{eq:electrostatic} accounts for all these interactions.
The electric field $\bfE(\bfx)$ and electric potential $\phi(\bfx)$ are related by the classical relation:
\begin{equation}\label{eq:E_elec_potential_relation}
    \bfE(\bfx)= - \nabla_{\bfx} \phi(\bfx).
\end{equation}

The external-like potential $w(\bfx, \bfu)$ acting on the dielectric polymer chain due to the electric field $\bfE(\bfx)$ is obtained as \cite{fredrickson2006equilibrium}:
\begin{equation}\label{eq:w_afo_E}
    w(\bfx, \bfu)
    =  - \frac{4 \pi V}{2 k_B T} \ \bfp_{seg}(\bfx,\bfu)\cdot\bfE(\bfx).
\end{equation}

\subsection{Polymer segment dipole response}\label{sec:monomer_response}
We consider that the dielectric response of chain segments is linear in the electric field\footnote
{
One can easily replace the linear dielectric response we used in this work with more general nonlinear responses.  
}.
However, to satisfy the rotation invariance of the monomers, this response is necessarily nonlinear in the orientation $\bfu$.
The polarization response of a polymer chain segment is defined as:
\begin{equation}\label{eq:polarization_linear_form}
	\bfp_{seg}(\bfx, \bfu):= \epsilon_0 \bfbeta(\bfu) \bfE(\bfx),
\end{equation}
where $\bfbeta$ is the molecular polarizability tensor of the chain segment that depends on the segment orientation $\bfu$. 
We assume the polarizability tensor $\bfbeta$ to be transversely isotropic as \cite{cohen2016electromechanical,cohen2016electroelasticity}:
\begin{equation}\label{eq:linear_polarizability}
	\bfbeta(\bfu)= \beta_{\parallel} \bfu \otimes \bfu + \beta_{\perp} \left(\bfI-\bfu \otimes \bfu\right),
\end{equation}
where $\beta_{\parallel}$ and $\beta_{\perp}$ are segment polarizabilities along the segment orientation and transverse to the segment orientation, respectively, and $\bfI$ is the identity tensor. 
The anisotropy in polymer dielectric response scales with the difference between $\beta_{\parallel}$ and $\beta_{\perp}$.


In summary of the theory, the electric potential $\phi$ provides the external potential $w(\bfx, \bfu)$ (see \eqref{eq:E_elec_potential_relation}-\eqref{eq:w_afo_E}) that connects to the partial partition functions $q$ and $q^*$ (see \eqref{eq:q_PDE}-\eqref{eq:q_star_PDE}). 
The quantities $q$ and $q^*$ are used to obtain the partition function $Q[w]$ and average segment density $\langle \hat{\rho}(\bfx, \bfu)  \rangle $ (see \eqref{eq:Q}, \eqref{eq:rho_avg}). 
The polarization $\bfp(\bfx)$ relates to $Q[w]$, $\langle \hat{\rho}(\bfx, \bfu)  \rangle $, and $\bfp_{seg}(\bfx, \bfu)$ (see \eqref{eq:avg_polarization_x}). Finally, $\bfp(\bfx)$ is used to obtain the electric potential $\phi$ through the electrostatic equation (see \eqref{eq:electrostatic}), closing the loop.

The theory presented here accounts for the entropic contributions via the partition function $Q[w]$.
Further, we account for the electrostatic energy of the dielectric polymer chain in an externally applied electric field through the external potential $w$. In general, by using the appropriate form of the functional dependence of $w(\bfx, \bfu)$ on external field and segment orientation $\bfu$, the presented theory can be directly extended for polymer chains influenced by a broad class of external (e.g., electric or magnetic) fields \cite{fredrickson2006equilibrium}. Specific examples include polymer chains with induced or permanent electric or magnetic dipoles along the chain backbone, as well as liquid crystalline polymers.

\section{Results and discussion}

We constrain the chain ends such that the chain orientation vector is along $\hat{\bfe}_2$ direction (Fig. \ref{fig:intro}). 
We induce chain stretch by varying the length of the chain orientation vector beyond its base value chosen as its root mean square average of $aN^{1/2}$ for a flexible polymer \cite{fredrickson2006equilibrium}.
We apply an electric field (denoted by $\tilde{\bfE}_0$ in a rescaled setting) on the domain boundary inside which the polymer chain is placed, resembling a realistic boundary condition.
We numerically solve the model 
(details in Sec. Numerical method) and compute the electric field $\bfE(\bfx)$,  polarization $\bfp(\bfx)$, and average segment density $\langle \hat{\rho}(\bfx, \bfu)  \rangle$ inside the domain. 
The rescaled quantities,  electric field $\tilde{\bfE}$,  polarization $\tilde{\bfp}$, and spatial average segment density $\tilde{\rho}$ are defined as:
\begin{equation}\label{eq:rescaling}
    \tilde{\bfE} = \frac{\bfE}{\sqrt{ \frac{2 k_B T}{L_c^3}}}, \quad \tilde{\bfp}=\frac{\bfp}{\sqrt{ \frac{2 k_B T}{L_c^3}}}, \quad \tilde{\rho}= \int \dm\bfu \langle \hat{\rho}(\bfx, \bfu)  \rangle.
\end{equation}
The rescaled critical value of the applied electric field for chain collapse is denoted by $|\tilde{\bfE}_c|$. 
We consider the dielectric response of chain monomers to be linear in electric field $\bfE$ 
but nonlinear in chain orientation $\bfu$, as given in \eqref{eq:polarization_linear_form}-\eqref{eq:linear_polarizability}.
We account for the anisotropy in the dielectric response of the polymer chain by choosing $\beta_{\parallel}=1$ and $\beta_{\perp}=0.5$ in \eqref{eq:linear_polarizability}, i.e., higher electrical polarizability along the polymer chain backbone orientation compared to the transverse direction. Such anisotropy is typical for real polymers having conjugated bond segments along the backbone \cite{schindler2006single, blythe2005electrical, jackson2015conformational}, liquid crystal elastomers (LCEs) \cite{spillmann2007anisotropic, fowler2021liquid, anglaret2005molecular, ware2015voxelated}, and biopolymers such as silk fibroin \cite{shi2014high, wu2022mesoscopic, chorsi2019piezoelectric, nepal2023hierarchically, notbohm2015microbuckling, song2024high, kim2010dissolvable}. 

First, we demonstrate chain collapse in high electric fields at different orientations of the applied electric field. 
Then, the effect of scaling of the dipole-dipole interactions on the equilibrium electric field and polarization is discussed. 
Finally, we present the variation of the critical electric field for chain collapse with the chain stretch and the orientation of the applied electric field.

\subsection{Chain collapse in a high electric field}
The chain collapse phenomenon is indicated by the sudden changes in the equilibrium properties of the constrained polymer chain observed using the model when the strength of the applied electric field increases beyond a critical value. 
Fig. \ref{fig:intro} shows the equilibrium properties of the chain in non-collapsed and collapsed states when the applied electric field 
$\tilde{\bfE}_0$ is aligned with the chain orientation.
When the strength of the applied electric field is less than the critical value for chain collapse $(|\tilde{\bfE}_0| < |\tilde{\bfE}_c|)$, the chain monomers are mainly distributed along the chain orientation (Fig. \ref{fig:intro}B). 
The corresponding electric field (Fig. \ref{fig:intro}C) and polarization (Fig. \ref{fig:intro}D) of the non-collapsed chain are observed to be largely concentrated near the chain ends. 
When the applied electric field is higher than the critical value $(|\tilde{\bfE}_0| \geq |\tilde{\bfE}_c|)$, the chain suddenly collapses.
 For the collapsed chain, the average segment density is observed to be significantly higher near the center location along the chain orientation vector where the chain collapses (Fig. \ref{fig:intro}F). This implies that a large number of polymer segments have accumulated near the chain collapse location. 
We observe that both the electric field (Fig. \ref{fig:intro}G) and polarization (Fig. \ref{fig:intro}H) of the collapsed chain are not only concentrated near the chain collapse location but are also orders of magnitude higher than in the non-collapsed state. 

We also notice the chain collapse phenomenon when we change the orientation of the applied electric field from being aligned to orthogonal to the chain orientation (see Supplementary Material Sec. 1 for details).
The sharp changes in density, electric field, and polarization signify that chain collapse is an instability.


\subsection{Effect of nonlocal dipole-dipole interactions}

\begin{figure*}[!t]
\centering
\includegraphics[width=1\textwidth]{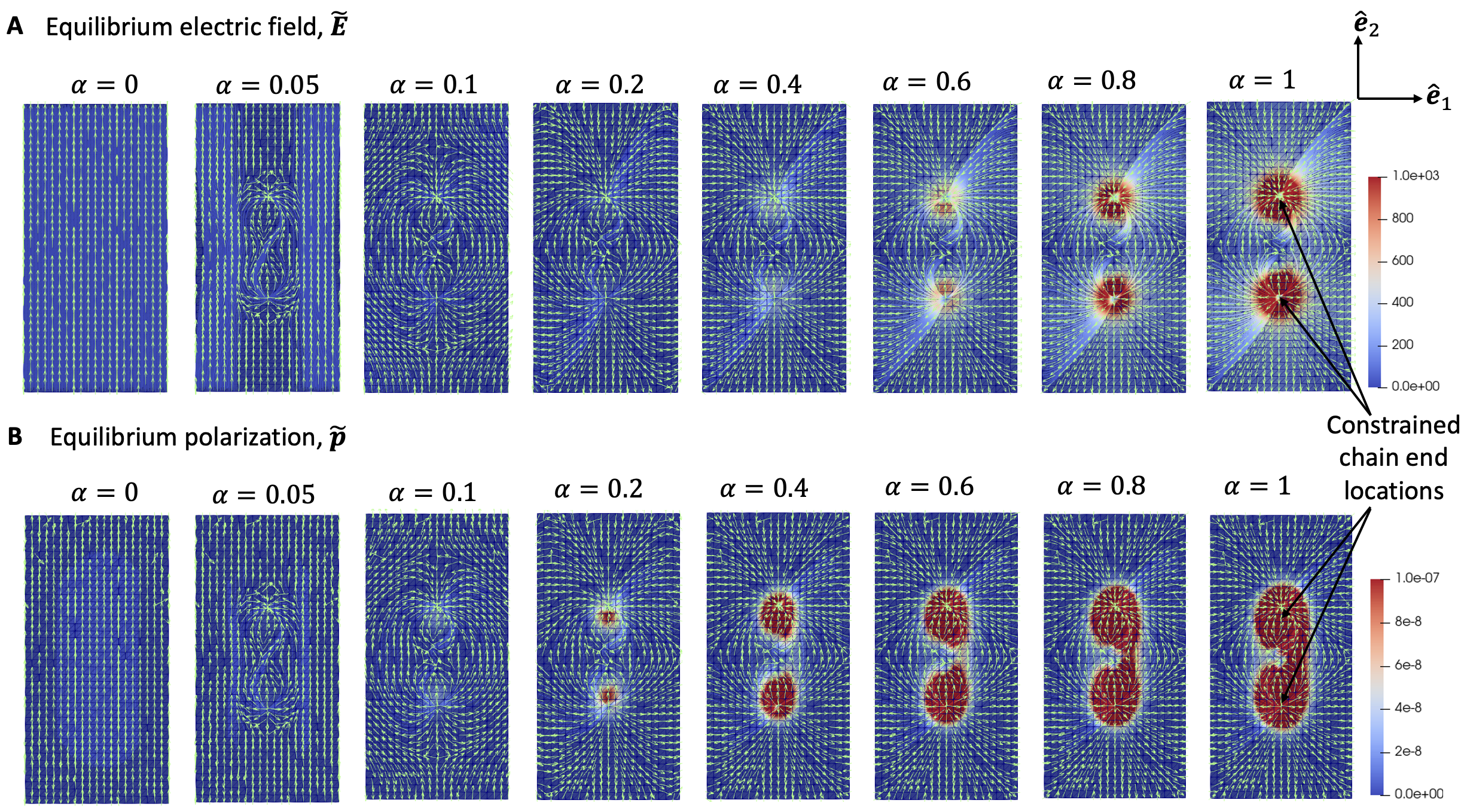}
\caption{Effect of dipole-dipole interactions on the equilibrium properties of the polymer chain when the applied electric field $(\tilde{\bfE}_0=1 \hat{\bfe}_2)$ is aligned with the chain orientation. 
A) Electric field $\tilde{\bfE}$ and B) polarization $\tilde{\bfp}$ for different values of $\alpha$ indicated above each panel in the figure. $\alpha=0$ and $\alpha=1$ represent the cases of no dipole-dipole interactions and full dipole-dipole interactions, respectively. The dipole-dipole interactions significantly change the distribution pattern for the electric field and average polarization at equilibrium. As we gradually increase $\alpha$, i.e., increasing the dipole-dipole interactions, we observe a gradual increase in the strength of the electric field and polarization near the constrained chain ends.}\label{fig:effect_of_dipole_interaction_on_E_p}
\end{figure*}

Importantly, we observe chain collapse only when nonlocal dipolar interactions are accounted for and when the polarization response of the monomer segments is anisotropic. 
To investigate the effect of nonlocal dipole-dipole interactions on the equilibrium properties of the dielectric polymer chain, we provided a way to vary the amount of dipole-dipole interactions being accounted for in the statistical field theory in \cite{khandagale2024statistical}. We achieved this by introducing a scalar parameter $\alpha \in [0,1]$ into the electrostatic equation
that connects the electric potential and the polarization for a dielectric material (see \eqref{eq:electrostatic}). $\alpha=0$ and $\alpha=1$ represent the two extreme cases of no dipole-dipole interactions and accounting for full dipole-dipole interactions, respectively.

Consider a case where an applied electric field is aligned with the chain orientation $(\tilde{\bfE}_0=1  \hat{\bfe}_2)$. 
As we gradually increase the value of $\alpha$, effectively increasing the dipole-dipole interactions being accounted for, we observe that the electric field becomes primarily concentrated near the constrained chain ends and forms a distribution pattern as shown in Fig. \ref{fig:effect_of_dipole_interaction_on_E_p}A.
Vector arrows of the electric field in Fig. \ref{fig:effect_of_dipole_interaction_on_E_p}A imply that polarization-induced electric charge separation is more dominant near the constrained chain ends, where the average segment density of the chain is higher (Fig. \ref{fig:intro}B). 
The higher segment density near the chain ends results in a higher dipole-dipole interaction effect, making the electric field largely concentrated near the constrained chain ends (Fig. \ref{fig:effect_of_dipole_interaction_on_E_p}A).


Similar to the electric field, the polarization (Fig. \ref{fig:effect_of_dipole_interaction_on_E_p}B)  also becomes largely concentrated near the constrained chain ends as we increase the dipole-dipole interactions being accounted for. 
The polarization in the domain (Fig. \ref{fig:effect_of_dipole_interaction_on_E_p}B) aligns with the local electric field (Fig. \ref{fig:effect_of_dipole_interaction_on_E_p}A). 
These observations are consistent with the linear polarization response of polymer segments in the electric field considered in this work. 
In the context of dielectric polymer networks, the constrained chain end locations would represent the polymer chain cross-link points where we suspect the effect of dipole-dipole interactions to be dominant.


\subsection{Effect of chain stretch on chain collapse}
\begin{figure}%
\centering
\includegraphics[width=\linewidth]{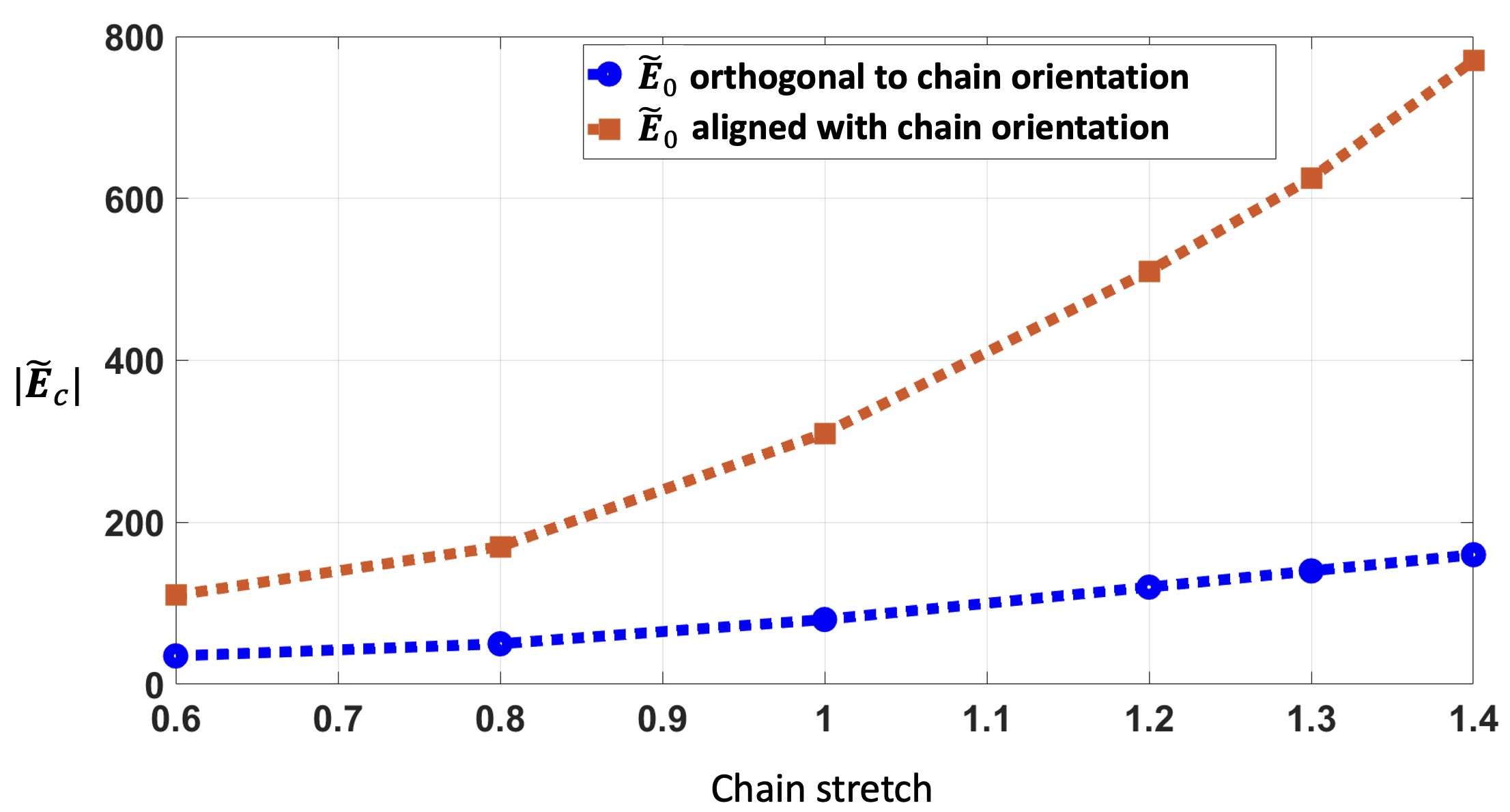}
\caption{Critical electric field values $(|\tilde{\bfE}_c|)$ for the occurrence of chain collapse for different chain stretches. Chain stretch is defined as the ratio of the length of the chain orientation vector and its root mean square average of $aN^{1/2}$ for the undeformed flexible polymer chain.
     }\label{fig:E_chain_collapse}
\end{figure}
The effect of mechanical deformation (chain stretching) on chain collapse is shown in Fig. \ref{fig:E_chain_collapse}. 
We stretch the chain by increasing the distance between the constrained chain ends. 
We observe that, for a given orientation of the applied electric field, the critical electric field value for chain collapse increases with the chain stretch. 
For a given chain stretch, the critical electric field for chain collapse is higher and more sensitive to the chain stretch when the applied field is aligned, as compared to orthogonal to the chain orientation. 
When the applied electric field is orthogonal to the chain orientation, the chain monomers would rearrange and rotate to align with the applied electric field orientation. 
We suspect that this rearrangement and mobilization of monomers induce chain collapse instability at a lower applied electric field when the applied field is orthogonal to the chain orientation. 
The stretch-dependent features of chain collapse offer new pathways to exploit chain collapse instability for deformation-sensitive rapid actuation and sensing using dielectric polymer-based soft matter.


\section{Concluding remarks}
Our findings reveal that the collapse of a dielectric polymer chain constrained at both ends in a high electric field is driven by the nonlocal dipole-dipole interactions and anisotropy in the polarization response of chain monomers, a characteristic of many real polymers. 
%
The chain collapse occurs sharply and leads to significant changes in the average segment density,  electric field, and polarization of the polymer chain. 
The electric field and polarization of the collapsed chain are orders of magnitude higher and primarily concentrated near the center location along the chain orientation vector, where the chain collapses. 
We have provided a novel approach to scale the nonlocal dipole-dipole interactions accounted for in the statistical field theory model for dielectric polymers. This approach enables us to gain deeper insights into the complex effects of nonlocal dipole-dipole interactions that are challenging to account for. 
The increase in the critical electric field for chain collapse with chain stretching, as reported here, is consistent with the experimentally observed trend for electrical breakdown strength with the stretching of dielectric elastomer \cite{trols2013stretch,huang2012thickness}. However, further investigation is needed to ensure that the chain collapse of a single dielectric polymer chain drives the electrical breakdown of the dielectric elastomers. 
A typical value of the rescaled critical electric field $|\tilde{\bfE}_c|=625$ corresponds (using \eqref{eq:rescaling}) to $1.79$kV/m, feasible to implement in experiments at room temperature ($298$K) for the collapse of a dielectric polymer chain having contour length of $0.1 \mu$m. 

The sharp collapse of the dielectric polymer chain presents new opportunities for rapid actuation and sensing of dielectric polymers using electrical stimuli. It provides a way to fabricate smart material structures having spatially varying multi-functional responses in an external electric field by embedding collapse-prone dielectric polymer chains in the material architecture. 
%
%
Accounting for the nonlinear polarization response of the segments with respect to the electric field, repulsive inter-segment interactions \cite{khandagale2023statistical}, and electrostatic effects of a medium surrounding the polymer chain, which could influence the chain collapse phenomenon, are important future research directions.
This work offers a way to explore the role of nonlocal dipole-dipole interactions in the dielectric polymer instabilities \cite{zurlo2017catastrophic, wang2011creasing, zhang2011mechanical, huang2012thickness, keplinger2010rontgen, niemeyer1984fractal, suo2010theory}. 
Extending the single-chain model to polymer networks \cite{khandagale2023statistical, purohit2011protein} provides an exciting opportunity to discover and study novel functional responses of dielectric elastomers to electric field stimuli. This will also help us to investigate whether chain collapse plays a role in experimentally observed instabilities and electrical breakdown of dielectric elastomers, which are major obstacles to their real-world applications.
%


\section{Numerical method}\label{sec:numerical_method}
\label{app:numerics}

The configuration space $(\bfx, \bfu)$ is 5-d, composed of segment spatial position $\bfx$ in 3-d and segment orientation vector $\bfu$ represented in 2-d (polar and azimuthal angles) spanning a unit sphere.
We numerically solve the model by reducing this 5-d configuration space to 3-d: we choose two spatial dimensions (i.e., $\bfx = (x_1, x_2) \in\Omega\subset \mathbb{R}^2$) and $\bfu = (\cos \phi , \sin \phi)$, where $\phi \in  [0, 2 \pi)$.
This enables us to develop a finite-element implementation that does not require periodicity, which would introduce spurious numerical artifacts.

For numerical implementation, the length scale in the problem is non-dimensionalized by $L_c$.
The computational domain is chosen to be $-0.1 \leq \tilde{x}_1 \leq 0.1, -0.2 \leq \tilde{x}_2 \leq 0.2 $, where $\Tilde{\bfx}=(\Tilde{x}_1, \Tilde{x}_2)= \left(\frac{x_1}{L_c}, \frac{x_2}{L_c} \right)$ is the nondimensional spatial coordinate.
In this work, we use persistence length $\lambda=L_c/1000$ to model a flexible polymer chain, and chain segment length $a=L_c/100$ (i.e., number of polymer segments in the chain to be $N=100$) for numerical computation. 
Since each polymer segment is formed by a set of consecutively connected monomers along the polymer backbone in a coarse-grained setting, and since monomers would be freely jointed along the chain backbone for a flexible polymer chain, one can expect the distance $\lambda$ at which orientational correlations decay to be less than the chain segment length $a$. This is consistent with our chosen parameter values for $\lambda$ and $a$ having relation as $\lambda=a/10$.

\begin{algorithm}[!t]
	\caption{Computing equilibrium properties of dielectric polymer chain using self-consistent iterative procedure}
	\begin{algorithmic}[1]
		\While{$\Delta Q > \epsilon = 10^{-3}$} 
		\State Compute $ \bfp(\bfx) = V \int \dm \bfu \
		\bfp_{seg}(\bfx, \bfu)  \langle  \hat{\rho}(\bfx, \bfu) \rangle $ \Comment{ $\bfp_{seg}(\bfx, \bfu)= \epsilon_0 \bfbeta(\bfu) \bfE(\bfx)$} 
		\State Solve for $\phi(\bfx)$: $\nabla_{\bfx}^2 \phi(\bfx)= \frac{\alpha}{\epsilon_0} \nabla_{\bfx} \cdot  \bfp(\bfx) $, $\quad$ given  $\phi(\bfx)= -\bfE_0\cdot \bfx$ on ${\partial \Omega}$
		\State Compute $\bfE(\bfx)=-\nabla_{\bfx} \phi(\bfx)$
		\State Compute $ w(\bfx, \bfu)= -  \frac{4 \pi V}{2 k_B T} \Big[ \epsilon_0 \bfbeta(\bfu) \bfE(\bfx) \Big] \cdot \bfE(\bfx)$
		\State Compute $q$ and $q^{*}$, by solving  \eqref{eq:q_PDE} and \eqref{eq:q_star_PDE}, respectively
		\State Compute $Q[w]$ and $\langle \hat{\rho}(\bfx, \bfu) \rangle$, using \eqref{eq:Q} and \eqref{eq:rho_avg}, respectively
		\EndWhile
		\State \textbf{Outputs:} Equilibrium quantities
        $ \bfE^{eq}(\bfx), Q^{eq}, \langle \hat{\rho}(\bfx, \bfu) \rangle^{eq}, \bfp^{eq}(\bfx) $
	\end{algorithmic}\label{scft_algorithm}
\end{algorithm}

The numerical iterative procedure to obtain the equilibrium properties of a dielectric polymer chain is shown in Algorithm \ref{scft_algorithm}. 
At the initial iteration step, we simply use $\phi (\bfx) = -\bfE_0 \cdot \bfx$ to obtain the electric field, and use it to compute $\bfp$ by following the procedure that we perform at every iteration, as summarized below. 
We use the electrostatic equation to continue the numerical iteration from step $n$ to step $n+1$; specifically, we use $\bfp^n(\bfx)$, the polarization at any iteration step $n$, to obtain the electric potential at the next iteration step, $\phi^{n+1}(\bfx)$ using:
\begin{equation}\label{eq:iteration}
    \nabla_{\bfx}^2 \phi^{n+1}(\bfx)=  \frac{\alpha}{\epsilon_0} \nabla_{\bfx} \cdot \bfp^n(\bfx), \quad \text{ given } \phi^{n+1}(\bfx)= -\bfE_0\cdot \bfx \text{ on } \partial\Omega.
\end{equation}  
The quantity $\phi^{n+1}(\bfx)$ is used to obtain $w^{n+1}(\bfx, \bfu)$ using \eqref{eq:E_elec_potential_relation} and \eqref{eq:w_afo_E}. 
Using $w^{n+1}(\bfx, \bfu)$, we again solve PDEs in \eqref{eq:q_PDE} and \eqref{eq:q_star_PDE} to obtain $q$ and $q^{*}$ and use them to compute $Q[w^{n+1}]$ and $\langle \hat{\rho}(\bfx, \bfu) \rangle^{n+1} $ using \eqref{eq:Q} and \eqref{eq:rho_avg}, respectively. 
Next, we compute $\bfp^{n+1}(\bfx)$ using \eqref{eq:avg_polarization_x_u_appendix} and \eqref{eq:avg_polarization_x}, which is used in \eqref{eq:iteration} to continue the iteration.
We continue the self-consistent iteration procedure until the energy term $-k_B T \log Q$ has converged, which we verify by checking the change in $Q$ across successive iterations. 
The converged quantities $\bfE(\bfx), Q[w], \langle \hat{\rho}(\bfx, \bfu) \rangle, \bfp(\bfx)$ are the thermodynamic equilibrium quantities for the polymer chain. We note that, although we introduced $w$ as an external-like potential acting on the polymer chain, the equilibrium potential $w$ at the end of self-consistent iterations accounts for the combined effects of internally generated interactions among the induced dipoles on the polymer segments as well as the local electric field.


We use the finite element method (see Supplementary Material Sec 2 for details) to solve the PDEs in \eqref{eq:q_PDE}, \eqref{eq:q_star_PDE}, and \eqref{eq:electrostatic} as used in \cite{ackerman2017finite}.
FEniCS, an open-source finite element method framework, is used for the numerical implementation 
\cite{logg2012automated}.

\section{Acknowledgments}\label{sec:acknowledgments}
\label{app:Acknowledgments}
We thank Carlos Garcia Cervera and Narayana Aluru for useful discussions.




\section{Funding}
We thank NSF (DMS 2108784, DMREF 1921857), BSF (2018183), and AFOSR (MURI FA9550-18-1-0095) for financial support; and NSF for XSEDE computing resources provided by Pittsburgh Supercomputing Center.

\section{Author Contributions Statement}

Pratik Khandagale: Conceptualization,  Methodology, Software, Formal analysis, Investigation, Writing – original draft preparation, Writing – review \& editing.
Gal deBotton: Conceptualization, Writing – review \& editing.
Timothy Breitzman: Conceptualization, Writing – review \& editing.
Carmel Majidi: Conceptualization, Writing – original draft preparation, Writing – review \& editing.
Kaushik Dayal: Conceptualization, Methodology, Formal analysis, Writing – original draft preparation, Writing – review \& editing.

\section{Competing Interest Statement}

There are no competing interests.



\section{Data availability}
A version of the code developed for this work is available at  \\
\url{https://github.com/pkhandag/polarizable-polymer.git}


\end{document}



\maketitle

\SItext


\section{Chain collapse when applied electric field is orthogonal to chain orientation}\label{app:chian_collapse_orthogonal}

\begin{figure*}[!t]
\centering
{\includegraphics[width=\textwidth]{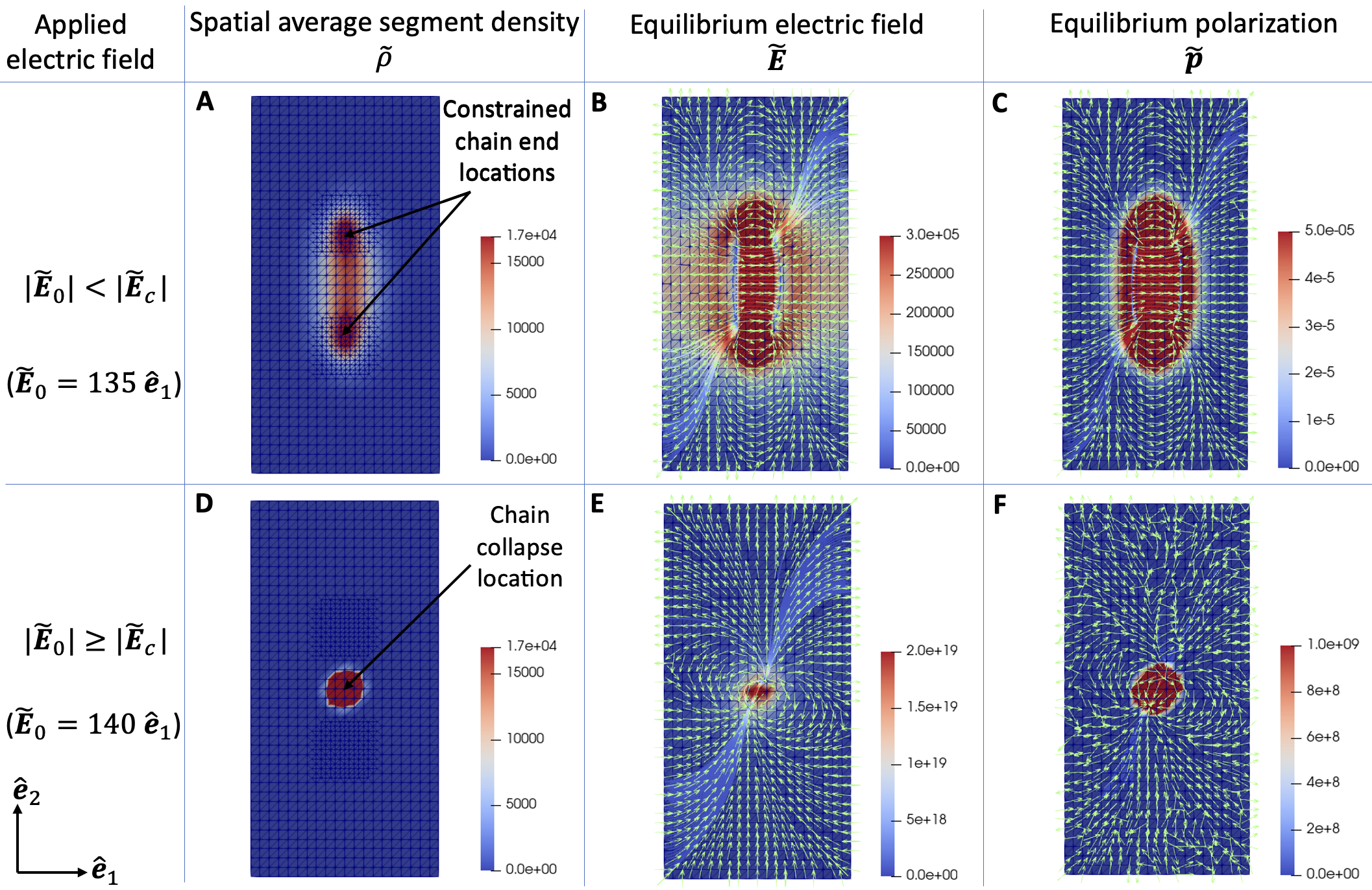}}
\caption{Chain collapse when the applied electric field is orthogonal to the chain orientation. A) Spatial average segment density $\tilde{\rho}$, B) electric field $\tilde{\bfE}$, and C)  polarization $\tilde{\bfp}$ for the non-collapsed chain (here $\tilde{\bfE}_0= 135 \hat{\bfe}_1$). 
D) Spatial average segment density $\tilde{\rho}$, E) electric field $\tilde{\bfE}$, and F) polarization $\tilde{\bfp}$ for the collapsed chain when the strength of the applied electric field (here $|\tilde{\bfE}_0|=140 \hat{\bfe}_1$) is higher than a critical value (i.e., $|\tilde{\bfE}_0| \geq \tilde{\bfE}_c$).
In the collapsed state, the electric field and polarization have significantly altered spatial distribution, and they are orders of magnitude higher when compared to the non-collapsed state. 
}\label{fig:chain_collapse_theta_0}
\end{figure*}

The equilibrium properties of the chain in non-collapsed and collapsed states when the applied electric field $\tilde{\bfE}_0$ is orthogonal to the chain orientation are shown in Fig. \ref{fig:chain_collapse_theta_0}.
When the strength of the applied electric field is less than the critical value for chain collapse $(|\tilde{\bfE}_0| < |\tilde{\bfE}_c|)$, the chain monomers are mainly distributed along the chain orientation (Fig. \ref{fig:chain_collapse_theta_0}A). 
The electric field (Fig. \ref{fig:chain_collapse_theta_0}B) and  polarization (Fig. \ref{fig:chain_collapse_theta_0}C) of the non-collapsed chain are observed to be distributed along the chain orientation. 
When the strength of the applied electric field is higher than the critical value $(|\tilde{\bfE}_0| \geq |\tilde{\bfE}_c|)$, the chain collapses suddenly.
 For the collapsed chain, the average segment density is observed to be significantly higher near the center location along the chain orientation vector where the chain collapses (Fig. \ref{fig:chain_collapse_theta_0}D). This implies that a large number of polymer segments have accumulated near the collapse location. 
We observe that both the electric field (Fig. \ref{fig:chain_collapse_theta_0}E) and polarization (Fig. \ref{fig:chain_collapse_theta_0}F) of the collapsed chain are not only concentrated near the chain collapse location but are also orders of magnitude higher than in the case of the non-collapsed chain.

\section{Finite element formulation}\label{sec:Finite Element Weak Formulation}


We work in 2 spatial dimensions (i.e., $\bfx = (x_1, x_2) \in \Omega \subset \mathbb{R}^2$) and restrict the unit orientation vector to the unit circle (i.e., it is represented as $\bfu=(\cos \phi, \sin \phi)$, where $\phi \in [0, 2 \pi]$). 
The configuration space in $(\bfx,\bfu)$ is 3-dimensional, enabling us to use standard finite element method (FEM) meshing and shape functions.
In terms of $q(x_1, x_2, \phi,s)$, we rewrite the partial differential equation (PDE) for $q$ as:
\begin{equation}\label{eq:q_PDE_weak_0}
	\frac{\partial q}{\partial s}= -w q - L_c \left( \cos{\phi} \frac{\partial q}{\partial x_1} + \sin{\phi} \frac{\partial q}{\partial x_2} \right) + \frac{L_c}{2 \lambda} \left( \frac{\partial^2 q}{\partial \phi^2} \right).
\end{equation}
The contour coordinate $s$ is treated as a time-like variable.
Derivatives concerning parameter $s$ along the chain contour in PDEs for $q$ and $q^*$ are approximated using a Crank-Nicolson finite difference method with $100$ steps for parameter $s$ along the chain contour.
In terms of the discretization of $s$, we write:
\begin{equation}\label{eq:q_PDE_weak_1}
\begin{split}
	& \frac{q^{i+1}-q^i}{\Delta s}= \frac{f^{i+1}+f^i}{2}, \\ 
    & \text{ with } f^i
	=
	-w q^i - L_c \left( \cos{\phi} \frac{\partial q^i}{\partial x_1} + \sin{\phi} \frac{\partial q^i}{\partial x_2} \right) + \frac{L_c}{2 \lambda} \left( \frac{\partial^2 q^i}{\partial \phi^2} \right),
\end{split}
\end{equation}
where the superscripts $i$ and $i+1$ represent the discretized quantities along $s$.

The domain in configuration space is discretized using 1-st order Lagrange family finite elements.
We use a mesh with $20\times40$ finite elements to discretize in $\bfx$ and $30$ finite elements to discretize in $\bfu$, which is sufficiently refined that the quantities of interest are independent of the mesh.
The spatial mesh is finer around the chain ends,  and the Dirac delta functions in initial conditions for PDEs for $q$ and $q^*$ are approximated as peaked Gaussians.
The mesh is uniform in the $\bfu$ discretization.

Following the usual FEM procedure, we, first, multiply \eqref{eq:q_PDE_weak_1} by a test function $v(x_1, x_2, \phi)$; second, integrate over $\bfx$ and $\bfu$; third, use integration-by-parts and the divergence theorem to convert the second derivatives $\frac{\partial^2 q}{\partial \phi^2}$ to a product of first derivatives; and, fourth, eliminate the boundary terms using the assumed homogeneous Neumann boundary condition in $(x_1, x_2,\phi)$ to get the FEM weak form:
\begin{equation}\label{eq:WLC_weak_form}
	\begin{split}
	& \int\limits_{\Tilde{\bfx},\phi} \Bigg(  q^{i+1}v  + \frac{\Delta s}{2} w q^{i+1}v 
		+ \frac{ \Delta s}{2} \cos \phi \frac{\partial q^{i+1}}{\partial \Tilde{x}_1} v
		 + \frac{ \Delta s}{2} \sin \phi \frac{\partial q^{i+1}}{\partial \Tilde{x}_2} v 
        	+ \frac{ L_c \Delta s}{4 \lambda} \frac{\partial q^{i+1}}{\partial \phi} \frac{\partial v}{\partial \phi}  \Bigg)  \\
            & =  
		\int\limits_{\Tilde{\bfx},\phi} \Bigg(  q^{i} v - \frac{\Delta s}{2} w q^{i}v 
		- \frac{ \Delta s}{2} \cos \phi \frac{\partial q^{i}}{\partial \Tilde{x}_1} v 	 - \frac{ \Delta s}{2} \sin \phi \frac{\partial q^{i}}{\partial \Tilde{x}_2} v
		- \frac{L_c \Delta s}{4 \lambda} \frac{\partial q^{i}}{\partial \phi} \frac{\partial v}{\partial \phi}  \Bigg).
	\end{split}
\end{equation} 

\subsection{Finite element mesh convergence} \label{sec:finite element mesh convergence}
\begin{figure*}[!t]
	\centering
    \includegraphics[scale=0.3]{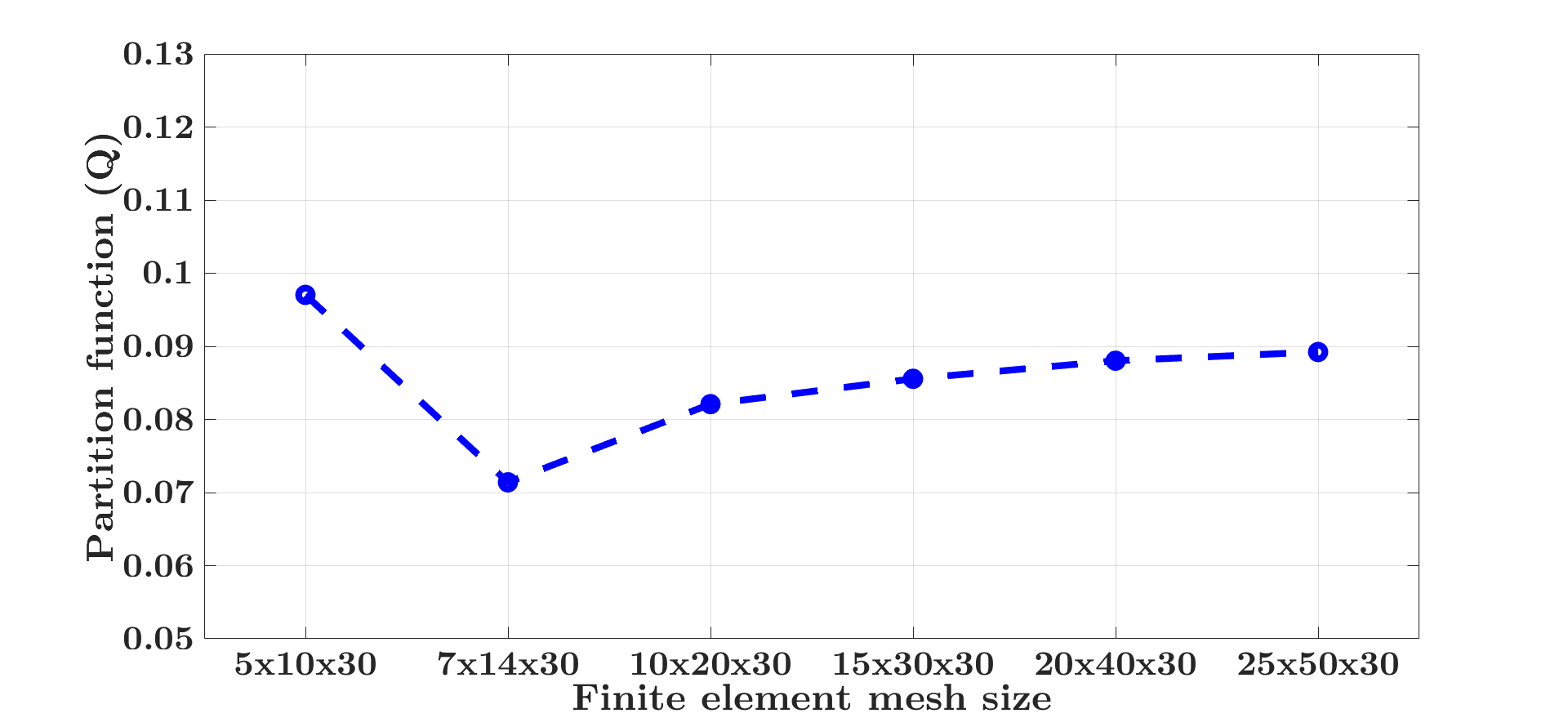}
    \caption{Finite element mesh convergence. A converged mesh size with $20$x$40$x$30$ number of finite elements (in $x_1, x_2, \phi$ dimension, respectively) is used for numerical computation. }
\label{fig:finite_element_mesh_convergence_check}
\end{figure*} 
Fig. \ref{fig:finite_element_mesh_convergence_check} shows a plot of the partition function of a worm-like chain for $\tilde{\bfE_0}=0.1 \hat{\bfe}_1$, $\alpha=1$, and length of chain orientation vector $L=1.3 aN^{1/2}$ for different mesh sizes. For numerical computation, we use a mesh size with $20 \times40\times30$ number of finite elements at which the partition function is converged.

\section{Scaling the electrical interactions between dipoles}
\label{sec:electro-appendix}

\subsection{Variational Structure}

The variational structure of our problem is to minimize a free energy $F[\bfp]$, subject to the constraint of electrostatics \cite{liu2013energy}:


\begin{align}
\label{eqn:energy-1}
     & F[\bfp] = 
        \int_{\bfx\in\Omega} W(\bfp) \dm\Omega_\bfx
        + \frac{\epsilon_0}{2} \int_{\bfx\in\Omega} |\nabla_\bfx \phi(\bfx)|^2 \dm\Omega_\bfx \nonumber \\
        & + \int_{\bfx\in\partial\Omega} \phi(\bfx) \left(-\epsilon_0 \nabla_\bfx \phi(\bfx) + \bfp(\bfx)\right)\cdot\hat\bfn \dm S_\bfx  ,  
    \\
    \label{eqn:constraint-1}
    & \qquad \text{ subject to }  -\epsilon_0 \nabla_\bfx^2 \phi (\bfx)= - \nabla_\bfx \cdot \bfp (\bfx) \text{ for } \bfx\in\Omega, \nonumber \\
    & \quad \text{ and } \phi(\bfx) = \phi_0(\bfx) \text{ prescribed on } \bfx\in\partial\Omega.
\end{align}

We use subscripts to indicate the variable with respect to which differentiation and integration are performed.
The potential $\phi$ is a prescribed function $\phi_0$ on the boundary $\partial\Omega$.

There are 3 contributions to the energy:
the first integral accounts for the energy required to change atomic and electronic configurations in the process of forming dipoles;
the second integral accounts for the energy stored in the electric field; 
and the last integral accounts for the energy due to boundary charges that are required to maintain the fixed boundary potential.
Note that with $\phi$ prescribed on the entire boundary, the electric field in free space outside $\Omega$ is constant and can be ignored.

Setting to zero the variation (i.e., functional derivative) of $F$ with respect to $\bfp$, while accounting for the constraint, gives the Euler-Lagrange equation \cite{marshall2014atomistic,yang2011completely}:
\begin{equation}
\label{eqn:EL-equation}
	\parderiv{W}{\bfp} +  \nabla_{\bfx} \phi = {\bf 0}.
\end{equation}
Notice, e.g., that if we use $W(\bfp) = \frac{1}{2 \epsilon_0} \bfp\cdot\bfbeta^{-1}\bfp$, we recover the response for a linear dielectric $\bfp = \epsilon_0 \bfbeta \bfE$, where $\bfbeta$ is the polarizability tensor.

\subsection{Electrostatic solution in the fixed far-field applied field ensemble}

The fixed far-field applied field ensemble corresponds simply to setting  $\phi_0(\bfx) = -\bfE_0\cdot\bfx$.
We decompose the electrostatic problem into two sub-problems: the first accounts for $\phi_0$ on $\partial\Omega$, and the second accounts for the dipole distribution in $\Omega$:
\begin{align}
    \label{eqn:phi0-defn}
    & \nabla_\bfx^2 \phi_0 (\bfx)= 0 \text{ on } \Omega, \quad \phi_0(\bfx) =  -\bfE_0\cdot\bfx \text{ on } \partial\Omega,
    \\
    \label{eqn:phi-tilde-defn}
    & \epsilon_0 \nabla_\bfx^2 \tilde\phi (\bfx)= \nabla_\bfx \cdot \bfp (\bfx) \text{ on } \Omega, \quad \tilde\phi(\bfx) = 0 \text{ on } \partial\Omega.
\end{align}
The first problem \eqref{eqn:phi0-defn} is trivial to solve: we have $\phi_0(\bfx) = -\bfE_0\cdot\bfx$ over the entire domain $\Omega$.
The second problem \eqref{eqn:phi-tilde-defn} can be solved formally in terms of the Green's function $G$ for the domain.
Using $\phi = \phi_0 + \tilde\phi$, we can write:
\begin{equation}
\label{eqn:phi-solution}
\begin{split}
    \phi(\bfx) 
    & 
    = \phi_0(\bfx) + \frac{1}{\epsilon_0} \int_{\bfy \in \Omega} G(\bfx,\bfy) \nabla_{\bfy} \cdot \bfp(\bfy) \dm\Omega_{\bfy}, \\
   & = \phi_0(\bfx) 
    	- \frac{1}{\epsilon_0} \int_{\bfy \in \Omega} \nabla_\bfy G(\bfx,\bfy) \cdot \bfp(\bfy) \dm\Omega_{\bfy},
    \\
    & \implies
    \nabla_\bfx \phi(\bfx) = -\bfE_0 - \frac{1}{\epsilon_0} \int_{\bfy \in \Omega} \bfK(\bfx,\bfy) \bfp(\bfy) \dm\Omega_{\bfy}.
\end{split}
\end{equation}
Here we have used that $G(\bfx,\bfy) = 0 \text { for } \bfy\in\partial\Omega$ from the symmetry of the Dirichlet Green's function \cite{jackson2021classical}.
We denote by $\bfK(\bfx,\bfy) := \nabla_\bfx \nabla_\bfy G(\bfx,\bfy)$ the symmetric matrix-valued dipole kernel operator; physically, this gives the electric field at $\bfx$ due to a dipole at $\bfy$.
This expression shows that $\bfE(\bfx)$ is a superposition of the far-field applied electric field and the field created by the dipoles, represented by the nonlocal integral expression.

\subsection{Approximations}

Eliminating the electrostatic constraint by using \eqref{eqn:phi-solution} in the Euler-Lagrange equation \eqref{eqn:EL-equation}, we have:
\begin{equation}
\label{eqn:EL-equation-2}
	\parderiv{W}{\bfp} -  \bfE_0 - \frac{1}{\epsilon_0} \int_{\bfy \in \Omega} \bfK(\bfx,\bfy) \bfp(\bfy) \dm\Omega_{\bfy} = {\bf 0}.
\end{equation}
From the perspective of \eqref{eqn:EL-equation-2}, \cite{cohen2016electroelasticity,grasinger2020statistical} and others completely neglect the nonlocal dipole-dipole term and solve:
\begin{equation}
    \parderiv{W}{\bfp}
    - \bfE_0
    = {\bf 0}.
\end{equation}
In \cite{grasinger2022statistical}, they discuss improving this by keeping the nonlocal term, but truncating it after near-neighbor interactions:
\begin{equation}
    \parderiv{W}{\bfp}
    - \bfE_0 
    - \left( \frac{1}{\epsilon_0} \int_{\bfy \in \Omega} \tilde\bfK(\bfx,\bfy) \bfp(\bfy) \dm\Omega_{\bfy}\right)
    ={\bf 0},
\end{equation}
where $\tilde\bfK$ is the truncated operator.
Importantly, $\tilde\bfK$ does not correspond to the kernel of a standard PDE, and hence it cannot be reformulated as a local PDE constraint.

\subsection{Scaling Dipole-Dipole Interactions v. Applied Field-Dipole Interactions}

In this paper, we consider the situation wherein the dipole-dipole term is scaled by a factor $\alpha$:
\begin{equation}
\label{eqn:EL-scaled}
    \parderiv{W}{\bfp}
    - \bfE_0
    - \alpha  \left( \frac{1}{\epsilon_0} \int_{\bfy \in \Omega} \bfK(\bfx,\bfy) \bfp(\bfy) \dm\Omega_{\bfy}\right)
    = {\bf 0}.
\end{equation}
We show below that this corresponds to using a linear combination of the far-field applied field and the fully-interacting electrostatic field as in \eqref{eqn:linear-mixing}.
First, we can see that \eqref{eqn:EL-scaled} is the Euler-Lagrange equation of \eqref{eqn:energy-1} with the electrostatic constraint \eqref{eqn:constraint-1} scaled as below:
\begin{equation}
	\label{eqn:scaled-interaction}
	\epsilon_0\nabla^2_\bfx \phi_\alpha = \alpha \nabla_\bfx \cdot \bfp \text{ on } \Omega, \quad \phi_\alpha = -\bfE_0\cdot\bfx \text{ on } \partial\Omega.
\end{equation}
Next, consider the case $\alpha=1$, corresponding to the dipoles fully interacting; denote the corresponding potential by $\phi_1$:
\begin{equation}
\label{eqn:full-interaction}
    \epsilon_0\nabla^2_\bfx \phi_1 = \nabla_\bfx \cdot \bfp \text{ on } \Omega, \quad \phi_1 = -\bfE_0\cdot\bfx \text{ on } \partial\Omega.
\end{equation}
Define the quantities $\tilde\phi_1 := \phi_1 + \bfE_0\cdot\bfx$ and $\tilde\phi_\alpha := \phi_\alpha + \bfE_0\cdot\bfx$.
These satisfy:
\begin{align}
    & \epsilon_0 \nabla^2_\bfx \tilde\phi_1 = \nabla_\bfx \cdot \bfp \text{ on } \Omega, \quad \tilde\phi_1 = 0 \text{ on } \partial\Omega,
    \\
    & \epsilon_0 \nabla^2_\bfx \tilde\phi_\alpha = \alpha \nabla_\bfx \cdot \bfp \text{ on } \Omega, \quad \tilde\phi_\alpha = 0 \text{ on } \partial\Omega.
\end{align}
Since these are linear problems with homogeneous BCs, we have that $\tilde\phi_\alpha=\alpha\tilde\phi_1$.
This implies:
\begin{equation}
\label{eqn:linear-mixing}
	\phi_\alpha = \alpha\phi_1 + (1-\alpha) \left(-\bfE_0\cdot\bfx\right) \implies
    \bfE_\alpha := - \nabla_\bfx\phi_\alpha = \alpha \bfE_1 + (1-\alpha) \bfE_0,
\end{equation}
where $\bfE_1 = -\nabla_\bfx \phi_1(\bfx)$ is the electric field in the fully-interacting setting.
Therefore, we have from \eqref{eqn:linear-mixing} that to study the scaled interaction case \eqref{eqn:EL-scaled}, we simply use a linear combination of the no-interaction potential $\phi_0 = -\bfE_0\cdot\bfx$ and the full-interaction potential $\phi_1$.














